\documentclass[aps, prl, twocolumn, showpacs, floatfix,10pt,superscriptaddress]{revtex4-2}
\usepackage{amsmath}
\usepackage{amsfonts}
\usepackage{amsbsy}
\usepackage{amssymb}
\usepackage{graphicx}
\usepackage{textcomp}
\usepackage[caption=false,singlelinecheck=false]{subfig}
\usepackage{xcolor}
\usepackage{mathrsfs}
\usepackage{mathtools}
\usepackage{bm}
\usepackage{gensymb}
\usepackage{braket,wasysym}
\usepackage[colorlinks,linkcolor=blue,citecolor=red,filecolor=magenta,urlcolor=red,breaklinks]{hyperref}
\usepackage[bottom]{footmisc}
\usepackage{verbatim}

\hypersetup{colorlinks=true, urlcolor=blue, citecolor=red, pdfborder={0 0 0}}
\usepackage{breakurl}
\usepackage{natbib}

\allowdisplaybreaks

\newcommand{\tcr}[1]{\textcolor{red}{#1}}

\normalfont
\def\be{\begin{equation}}
	\def\ee{\end{equation}}
\def\bea{\begin{eqnarray}}
	\def\eea{\end{eqnarray}}

\begin{document}
\title{
 Interaction tuned pattern-selective superconductivity:\\ Application to the dodecagonal quasicrystal
}

\author{Junmo Jeon}
\email{junmo1996@kaist.ac.kr}
\affiliation{Korea Advanced Institute of Science and  Technology, Daejeon 34141, South Korea}
\author{SungBin Lee}
\email{sungbin@kaist.ac.kr}
\affiliation{Korea Advanced Institute of Science and  Technology, Daejeon 34141, South Korea}

\date{\today}
\begin{abstract}
Quasicrystals exhibit superconductivity under the unique interplay of long-range order and strong inhomogeneity, distinguishing them from both crystalline and amorphous systems. Understanding how this structural complexity affects superconducting states and phase transitions remains an important open question. Here, we unveil anomalous superconductivity in a dodecagonal quasicrystal using the attractive Hubbard model within the Bogoliubov-de Gennes framework. We show that both the gap structure and critical temperature depend on the local pattern due to inhomogeneous charge distribution and kinetic terms. This leads to unconventional phase transitions between mixed phases where superconducting and normal metal regions coexist, even without external fields, which we term pattern-selective superconductivity. Furthermore, superconductivity can be anomalously suppressed even under stronger attractive interactions due to the alignment of the Fermi level with a spectral gap in the fragmented Hartree-shifted spectrum. These findings, not observed in conventional crystals, amorphous systems, or previous studies of quasicrystal superconductivity, highlight the distinct role of quasicrystalline order.
\end{abstract}
\maketitle

\textit{\tcr{Introduction---}} Superconductivity in quasicrystals provides a unique platform to explore the effects of aperiodicity on electronic order, distinct from both crystalline and amorphous systems\cite{tinkham2004introduction,kamiya2018discovery,uri2023superconductivity,takemori2024superconducting,sordelet1997quasicrystals,macia2020quasicrystals,suck2013quasicrystals,sun2024enhancement}. Unlike periodic crystals, quasicrystals exhibit long-range order without translational symmetry, giving rise to complex electronic structures with unconventional properties\cite{duncan2024critical,naka2005critical,poon1992electronic,kohmoto1987critical,jeon2021topological,jagannathan2021fibonacci}. This sets them apart not only from conventional superconductors but also from systems with weak quasiperiodicity, such as twisted multilayer systems, where moiré patterns introduce moderate inhomogeneity\cite{xia2025superconductivity,wu2018theory,chen2019signatures,torma2022superconductivity,uri2023superconductivity}. Strong quasiperiodicity in quasicrystals leads to highly inhomogeneous charge distributions, which can fundamentally alter the nature of Cooper pairing and critical behavior\cite{emery2000charge,ghosal2001inhomogeneous}. While superconductivity in twisted multilayer graphene and other weakly quasiperiodic systems is expected to be understood as a correlation effect within the framework of moiré band theory and Brillouin zone folding method\cite{yoshii2025brillouin,uri2023superconductivity,balents2020superconductivity,chen2019signatures}, the deeper structural complexity of quasicrystals introduces additional challenges that remain unresolved.

A key distinction in quasicrystals arises from the coexistence of extended, localized, and critical which is power-law scaling electronic states due to the lack of translational symmetry\cite{kohmoto1987critical,mace2017critical}. This results in a fragmented single-particle spectrum with multiple spectral gaps, complicating the formation of a global superconducting state and challenging the applicability of conventional BCS theory\cite{tinkham2004introduction,kamiya2018discovery,jagannathan2021fibonacci,takemori2024superconducting,takemori2020physical,tokumoto2024superconductivity}. The inhomogeneous charge distribution further induces a non-uniform Hartree shift, creating a spatially varying effective potential that directly impacts the pairing mechanism\cite{ghosal2001inhomogeneous}. These structural and electronic complexities suggest that superconductivity in quasicrystals could exhibit new phenomena beyond the gap structure itself, influencing not only the superconducting order parameter but also the critical temperature and phase transitions\cite{liu2024nematic,verbin2013observation,ghadimi2021topological,fan2021enhanced}.

Despite theoretical interest and recent experimental findings on superconductivity in quasicrystals, such as the Ta$_{1.6}$Te dodecagonal quasicrystal\cite{cain2020layer}, where deviations from BCS behavior in the heat capacity and its anomalous field response have been reported\cite{tokumoto2024superconductivity,terashima2024anomalous} — how strong quasiperiodicity shapes superconducting properties remains an open question. 
While previous studies have explored the effects of quasicrystal tiling on inhomogeneous gaps and Fulde–Ferrell–Larkin–Ovchinnikov states\cite{araujo2019conventional,sakai2017superconductivity,sakai2019exotic,ghadimi2024quasiperiodic}, the impact on superconducting phase transitions and the role of local structural patterns in determining critical temperatures remain largely unexplored. More broadly, the influence of quasiperiodicity on superconducting properties beyond pairing strength is still not well understood.

In this Letter, we propose two general mechanisms of superconductivity in quasiperiodic systems — pattern-selective and interaction-tuned — using the attractive Hubbard model within the Bogoliubov-de Gennes framework, with a dodecagonal quasicrystal as a specific example. Pattern-selective superconductivity arises from inhomogeneous charge distribution and kinetic energy, driven by variations in coordination number and hopping strength. This creates a spatially varying gap and local critical temperatures, leading to mixed phases where superconducting and metallic regions coexist even without external fields or unconventional pairing channels. By analyzing the heat capacity, we explore anomalous phase transitions of these mixed phases. While, interaction-tuned superconductivity stems from the fragmented nature of the spectrum in quasiperiodic system. The inhomogeneous Hartree shift can align the Fermi level with a spectral gap of electrons, causing anomalous suppression of superconductivity even for stronger interaction. Our findings show that strong quasiperiodicity and inhomogeneity not only reshape the superconducting gap structure but also drive unconventional critical phenomena, providing a unified framework for understanding superconductivity in complex quasiperiodic systems.

\begin{figure}[]
\centering
\includegraphics[width=0.5\textwidth]{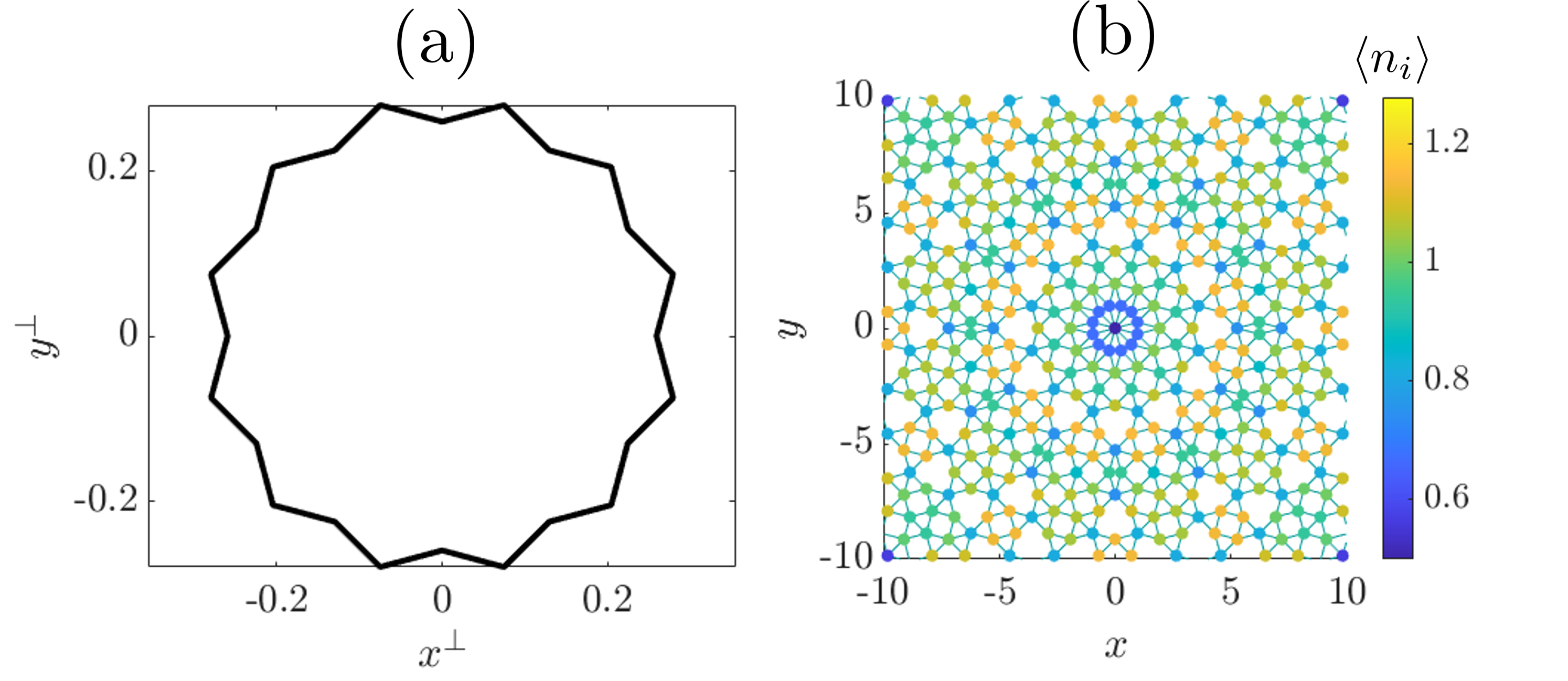}
\caption{\label{fig: ddQC} A dodecagonal quasicrystal constructred by the cut-and-project scheme. (a) Dodecagonal star-shaped window in orthogonal complement space. (b) Dodecagonal quasicrystal along with inhomogeneous charge density distribution. Here, lattice sites are shown as colored dots, with the color representing the charge density of electrons. For charge density, we consider $U/t=1$ and $\mu=-0.25t$. The number of sites is 5437.}
\end{figure}

\textit{\tcr{Attractive Hubbard model on the dodecagonal quasicrystal---}} Let us consider the dodecagonal quasicrystal constructed by the cut-and-project scheme (CPS)\cite{kellendonk2015mathematics}. The quasicrystalline lattice points are projected from a 4D hypercubic lattice, where the projection map onto a physical 2D plane is given by
\begin{align}
\label{projection_ddQC}
M = \frac{1}{a\sqrt{2}}
\begin{bmatrix}
\cos\left(\frac{\pi}{6}\right) & 1 & 0 & \cos\left(\frac{2\pi}{3}\right) \\
-\sin\left(\frac{\pi}{6}\right) & 0 & 1 & \sin\left(\frac{2\pi}{3}\right)
\end{bmatrix}
\end{align}
where $a=1.366$\cite{yamada2022four}. In the CPS, we consider the compact window $K$ in the orthogonal complement space, and select the hyperlattice points whose orthogonal complement projection images fall within $K$\cite{kellendonk2015mathematics}. By choosing $K$ as the dodecagonal shape, the resulting quasicrystalline lattice points can have the dodecagonal symmetry. In this paper, we adopt a dodecagonal star-shaped window shown in Fig.\ref{fig: ddQC} (a), a common choice in the study of dodecagonal quasicrystalline materials, such as Ba-Ti-O, Ni-Cr and V-Si-Ni\cite{yamada2022four,gahler1988crystallography}. Fig.\ref{fig: ddQC} (b) illustrates the dodecagonal quasicrystal. The sites whose distance is less than 1 is connected by the line. Note that the quasicrystal in Fig.\ref{fig: ddQC} (b) has not only squares and triangles, but also a deformed hexagon, known as the shield\cite{gahler1988crystallography,supple}. 

Now let us consider the attractive on-site Hubbard Hamiltonian given by
\begin{align}
    \label{H_hubbard}
    &H=-\sum_{i,j,\sigma}t_{ij}c_{i\sigma}^\dagger c_{j\sigma}-\mu \sum_i n_i-U\sum_i n_{i\uparrow}n_{i\downarrow},
\end{align}
where $t_{ij}$ is the hopping integral between sites $i$ and $j$, given by $t_{ij}=te^{1-R_{ij}}$ for $0<R_{ij}\le 1$ and zero otherwise, with $R_{ij}$ being the distance between the sites and $t=1$. Such distance-dependent hopping model enhances local kinetic diversity in systems with various neighboring atomic distances like quasicrystals. Our results can be generalized to other forms of hopping integrals. $\mu$ is the constant chemical potential maintained by the external reservoir. We set $\mu=-0.25t$, which corresponds to a filling fraction of about 0.4 for the non-interacting Hamiltonian, however, our results can be generalized to different $\mu$. $U>0$ is uniform on-site attractive interaction. We focus on the s-wave pairing only by neglecting off-site interactions and any other orbital degree of freedom. The nature of the Hamiltonian, Eq.\eqref{H_hubbard} is classified in terms of the dimensionless parameter, $U/t$. When $U/t$ is small enough, one can apply the Bogoliubov-de-Gennes (BdG) method to the mean field Hamiltonian given by
\begin{align}
    \label{H_bdg}
&H_{\mathrm{MF}}=-\sum_{i,j}t_{ij}c_{i\sigma}^\dagger c_{j\sigma}-\sum_i \tilde{\mu}_i n_i+\sum_i (\Delta_i c_{i\uparrow}^\dagger c_{i\downarrow}^\dagger+h.c.).
\end{align}
$\Delta_i=U\langle c_{i\uparrow}c_{i\downarrow}\rangle$ and $\langle n_i \rangle =\sum_{\sigma=\uparrow,\downarrow}\langle c_{i\sigma}^\dagger c_{i\sigma}\rangle$ are mean fields\cite{zhu2016bogoliubov}. $\Delta_i$ is known as the local superconducting order parameter. Specifically, the (local) superconducting phase is defined by non-zero $\Delta_i$. In the mean-field Hamiltonian, one should consider the Hartree shift of the local chemical potential. Without external magnetic field, this is given by $\tilde{\mu}_i=\mu+U \langle n_i \rangle /2$. The Hartree shift is important in inhomogeneous systems like quasicrystals due to their highly non-uniform charge density distribution. With an attractive interaction, charges tend to gather together, enhancing the spatial inhomogeneity of the Hartree shift.  Fig.\ref{fig: ddQC} (b) shows the inhomogeneous but pattern-dependent charge density distribution of electron for $U/t=1$. This influences on the single particle spectrum and wavefunctions. Thus, $\Delta_i$ and even critical temperature could be inhomogeneous through the system as we will show.

\begin{figure*}[]
\centering
\includegraphics[width=0.8\textwidth]{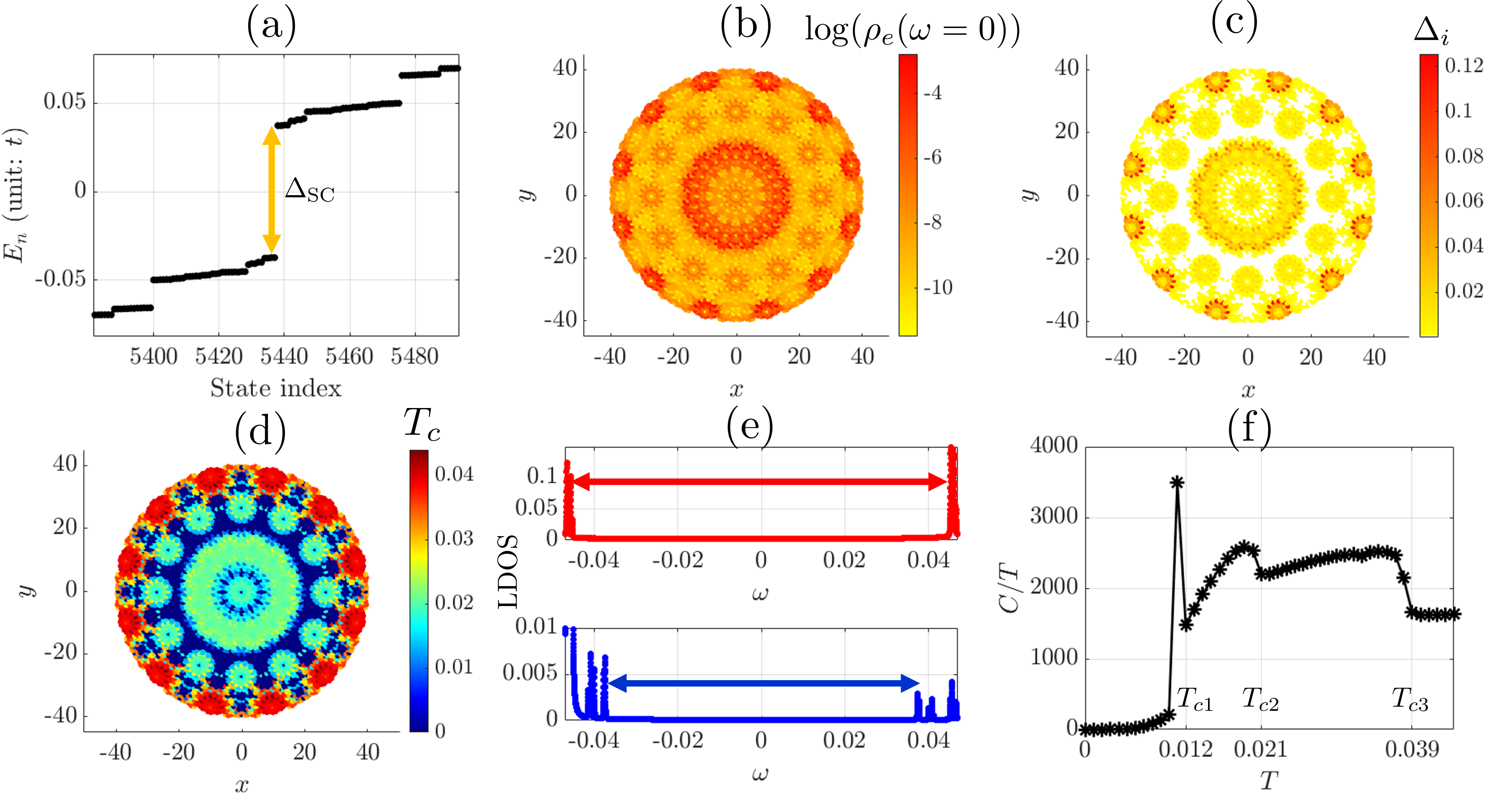}
\caption{\label{fig: nDLDOS} Pattern-selective superconductivity. (a) Spectrum of BdG energies, $E_n$ at zero temperature. The orange arrow is drawn to emphasize the superconducting gap, $\Delta_{\mathrm{SC}}$ of the spectrum. (b) The local density of states of electron near the Fermi level, considering the Hartree shifted effective single particle Hamiltonian. (c) Spatial distribution of order parameter, $\Delta_i$ at zero temperature. The white color represents normal metal region. (d) Spatially-dependent critical temperatures, $T_c$ whose pattern is similar to the order parameter. (e) Local density of states (LDOS) for two different sites with distinct critical temperatures. The blue curve represents the case of lower critical temperature. The red and blue arrows are drawn for emphasizing the width between first nontrivial resonant peaks of the LDOS for each case. (f) Heat capacity as the function of temperature. The drops of the heat capacity at three different temperatures, $T_{c1},T_{c2}$ and $T_{c3}$ indicate successive phase transitions to the normal metal. Before $T_{c3}$, the system is in the mixed phase. $\mu=-0.25t$ and $U=t$. The number of sites is $5437$. The temperatures, energy, $\omega$ and $\Delta_i$ are given by unit of $t$.}
\end{figure*}

The mean-field Hamiltonian $H_{\mathrm{MF}}$ can be solved by the BdG method\cite{zhu2016bogoliubov}. In detail, we consider the Bogoliubov transformation given by $c_{i\sigma}=\sum'_n[u^n_{i\sigma}\gamma_n-\sigma v^{n*}_{i\sigma}\gamma_n^\dagger]$, where $\sigma=\pm 1$ for $\uparrow,\downarrow$, respectively. $\gamma_n  (\gamma_n^\dagger)$ is annihilation (creation) operator of the Bogoliubov quasiparticle for the state indexed by $n$. The prime sign over the summation indicates that we only count states which
have a positive energy. Then, $H_{\mathrm{MF}}$ is diagonalized by this canonical transformation as $H_{\mathrm{MF}}=\sum_n E_n\gamma_n^\dagger\gamma_n+\mathrm{const}$. The energy eigenvalues $E_n$ can be obtained by solving the real space BdG equations given by
\begin{align}
    \label{BdG equations1}
    &E_nu^n_{i\uparrow}=\sum_j \mathcal{H}_{i\uparrow,j\uparrow}u^n_{j\uparrow}+\Delta_i v^n_{i\downarrow}\\ \nonumber
    &E_n v^n_{i\downarrow}=-\sum_{j}\mathcal{H}_{j\downarrow,i\downarrow}v^n_{j\downarrow}+\Delta_i^*u^n_{i\uparrow}
\end{align}
Here, $\mathcal{H}_{i\sigma,j\sigma}=-t_{ij}-\tilde{\mu}_i\delta_{ij}$ is the effective single particle Hamiltonian of electron. The eigenvalues, $E_n$ are symmetric around zero due to particle-hole symmetry (see Fig.\ref{fig: nDLDOS} (a)). At temperature $T$, we obtain the mean fields from the self-consistent gap equations given by
\begin{align}
    \label{gap}
    &\Delta_i=\frac{U}{2}\sum_n u^n_{i\uparrow}v^{n*}_{i\downarrow}\tanh{\left(\frac{E_n}{2T}\right)}
\end{align}
and
\begin{align}
    \label{gap}
    &\braket{n_i}=2\braket{n_{i\uparrow}}=2\sum_n\vert u^n_{i\uparrow}\vert^2 f(E_n),
\end{align}
where $f(E_n)=(1+e^{E_n/T})^{-1}$ is Fermi-Dirac distribution. We use $\hbar=k_B=1$.

\textit{\tcr{Pattern-selective superconductivity---}} We now focus on the anomalous features of superconductivity emergent in the dodecagonal quasicrystal. In quasicrystals, the exotic long-range ordered structure is given in terms of repetitive local patterns. It is quite natural to expect that the electronic structures and superconducting order parameters follow such local patterns. However, we assert that not only the superconducting gap structure but also the critical temperature locally varies due to the exotic patterns of the quasicrystal. This leads to multiple mixed phases with coexisting superconducting and normal metal regions, even without an external field—referred to as \textit{pattern-selective superconductivity}.

Before delving into details, we point out the importance of Hartree shift in quasicrystals. Since the electronic structure and charge distribution are non-uniform in quasicrystals due to their local patterns\cite{macia2020quasicrystals}, the Hartree shift becomes inhomogeneous. This leads to the pattern-dependent on-site potential that significantly influences on both spectrum and wavefunctions of effective electron Hamiltonian, $\mathcal{H}$. Thus, despite constant $\mu$, the electronic states near the Fermi level exhibit different characteristics as $U$ varies. Moreover, due to the aperiodic structure of $\mathcal{H}$, it has multiple spectral gaps, making variations in $U$ more significant in quasicrystals than in periodic systems. For instance, certain values of $U/t$ can place the Fermi level within a gap in the single-particle spectrum of $\mathcal{H}$, as discussed later.

To illustrate our claim, we use $U/t=1$ as an example, where the average size of the Cooper pair is much smaller than the system size, making the finite-size effect negligible\cite{sakai2017superconductivity,supple}. However, we emphasize that our results apply to a wide range of parameters as we will show. Fig.\ref{fig: nDLDOS} (b) exhibits the local density of states of electron, $\rho_e(i,\omega=0)$ of effective electron Hamiltonian $\mathcal{H}$ that is given by,
\begin{align}
    \label{electronldos}
    &\rho_e(i,\omega=0)=-\frac{1}{\pi}\mathrm{Im}(G^r(i,i,\omega=0)).
\end{align}
Here, $G^r(i,i,\omega)$ is the diagonal element of the retarded Green's function of $\mathcal{H}$, which is given by $(\omega-\mathcal{H}+i0^+)^{-1}$. $\mathrm{Im}(x)$ is the imaginary part of $x$.
Notably, the order parameter, $\Delta_i$ exhibits dodecagonal pattern that roughly follows $\rho_e(i,0)$ (compare Fig.\ref{fig: nDLDOS} (b) and (c)).

Interestingly, the critical temperatures ($T_c$) at which the local order parameter vanishes are also dependent on the surrounding pattern\cite{supple}. Therefore, it is essential to consider the concept of a local critical temperature when analyzing the system.
Fig.\ref{fig: nDLDOS} (d) exhibits spatial distribution of local $T_c$ for $U/t=1$. We emphasize that the local critical temperature arises in the bulk, not as a simple boundary effect.
The local critical temperature stems from both inhomogeneous $\rho_e$ and the competition between interaction and electron kinetic energy, which changes with the local pattern.
Notably, such non-uniform critical temperature is distinct from the case of disordered system, where the distribution of the local critical temperatures is random\cite{ghosal2001inhomogeneous}. In contrast, the critical temperature in the dodecagonal quasicrystal is spatially non-uniform but roughly follows the pattern of $\Delta_i$. Thus, the critical temperature distribution is neither spatially uniform nor random, highlighting a key difference between quasicrystals and conventional crystals or amorphous systems. Note that the local critical temperature could be captured from local density of states, LDOS at $T=0$, given by $\mathrm{LDOS}(i,\omega)=\sum_n\vert u_{i\uparrow}^n\vert^2\delta(E_n-\omega)$\cite{zhu2016bogoliubov}. Specifically, Fig.\ref{fig: nDLDOS} (e) shows LDOS as a function of $\omega$ for two sites with different local patterns. The wider gap in top panel indicates a higher local critical temperature.

The pattern-dependent critical temperature is intriguing, as it leads to mixed phases where superconducting and normal metal regions coexist even at zero magnetic field. Note that such mixed phases are different from the type-II superconductors which require a critical field\cite{collings2013applied,sharma2015type,tinkham2004introduction}. To show this, we investigate heat capacity, $C$ as the function of temperature. In detail, $C=T\frac{\partial S(T)}{\partial T}$, where $S(T)$ is the entropy at temperature $T$ given by
\begin{align}
    \label{entropy}
    S(T)=-\sum_n&[f(E_n)\log(f(E_n))\\ \nonumber &+(1-f(E_n))\log(1-f(E_n))].
\end{align}
In the normal metal regime, $C\propto T$, while in the uniform superconducting regime $C\propto e^{-\Delta(0)/T}$, where $\Delta(0)$ is the zero temperature superconducting gap. Thus, in conventional superconductivity, the heat capacity drops sharply when the phase transition from superconductor to normal metal occurs\cite{tinkham2004introduction}.

Fig.\ref{fig: nDLDOS} (f), however, shows successive drops of the heat capacity as the function of temperature. Comparing Fig.\ref{fig: nDLDOS} (d), these drops in heat capacity reflect the disappearance of the local order parameter. This indicates the presence of mixed phases even without external fields or unconventional p- or d-wave pairing channels\cite{zeng2016generalized,cao2020kohn,sakai2019exotic,li1993mixed,o1995s}. Such mixed phases are originated from the exotic tiling pattern of the dodecagonal quasicrystal and inhomogeneous charge distribution. Thus, we refer to these mixed phases as pattern-selective superconductivity.

Note that the second and third phase transitions are relatively gradual compared to the sharper first drop. This implies that the second and third phase transitions are beyond simple BCS theory prediction. Recent experiments on the Ta$_{1.6}$Te van der Waals dodecagonal quasicrystal have revealed similar deviations from the BCS theory prediction in its heat capacity\cite{tokumoto2024superconductivity}. Hence, our findings would support recent superconducting experimental results in dodecagonal quasicrystals.

\textit{\tcr{Interaction-tuned superconductivity---}} 
The exotic structure of the dodecagonal quasicrystal not only leads to a pattern-dependent critical temperature but also causes an intriguing $U$-dependence in superconductivity. In conventional systems, stronger attractive interactions enhance Cooper pairing and superconductivity. However, in quasicrystals, the fragmented single-particle spectrum with Hartree shifts results in irregular superconducting behavior as the attractive interaction strength changes, as we will demonstrate below.

\begin{figure}[]
\centering
\includegraphics[width=0.5\textwidth]{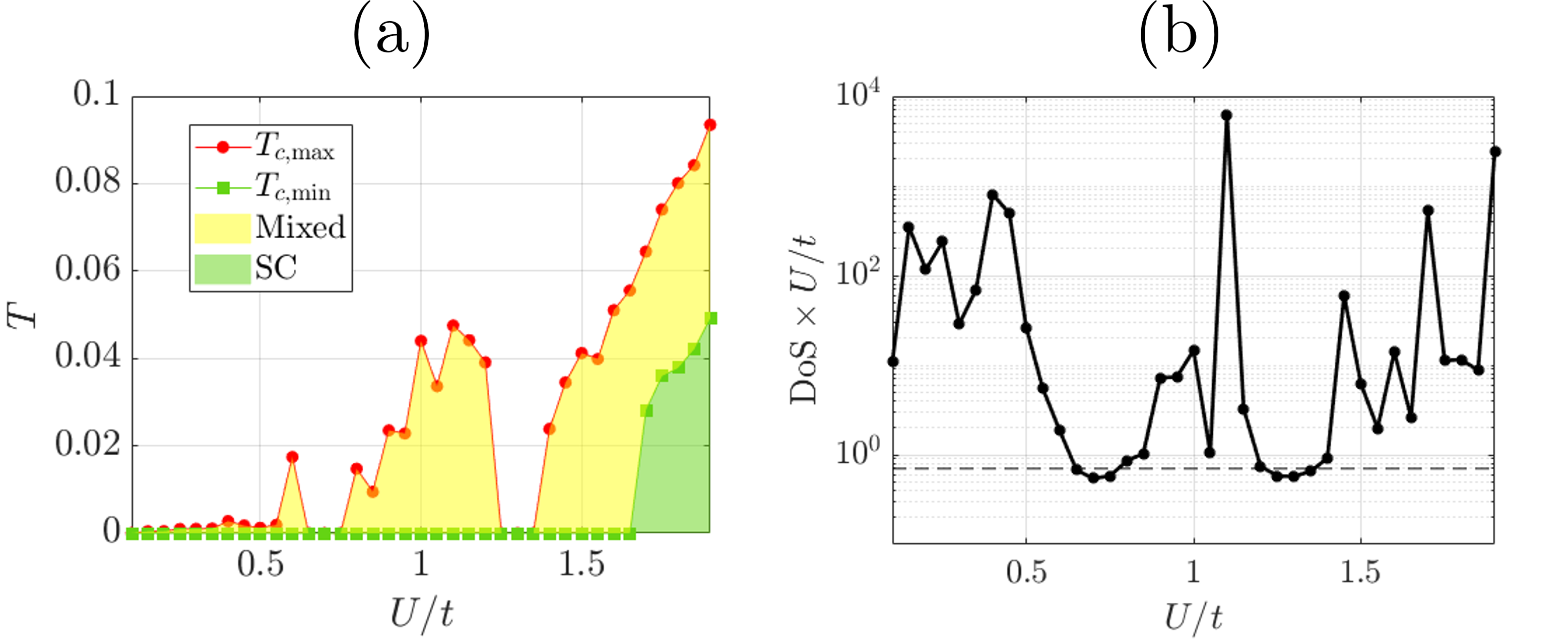}
\caption{\label{fig: dome} (a) Anomalous $U$-dependent superconductivity. The red circle and green square curves represent the maximum and minimum critical temperature as the functions of $U/t$. In the green area, the system is entirely superconducting. While, in the yellow area, the system is in the mixed phase where the superconducting region and normal metallic region coexist. The superconductivity does not appear even at zero temperature for $0.7\le U/t\le 0.8$ or $1.25\le U/t\le 1.35$. (b) Density of states (DoS) muliplied by $U/t$ of the single particle states at the Fermi level as the function of attractive interaction strength. The DoS anomalously drops for the same values of $U/t$ where the superconductivity disappears. $\mu=-0.25t$ and the number of sites is 5437.}
\end{figure}
To demonstrate $U$-dependence of superconductivity, we investigate the pattern-dependent critical temperatures as the function of $U/t$, which can be controlled by applying strain, for instance. Fig.\ref{fig: dome} (a) shows maximum (red circle curve) and minimum (green square curve) local $T_c$ as a function of $U/t$. As $U/t$ increases, we observe that the maximum critical temperature would decrease, and reaches zero at some values of $U/t$. Thus, the superconductivity would be absent even for stronger $U/t$. Such disappearance of superconductivity, even for stronger attractive interactions, is a highly unconventional phenomenon. We claim that this anomaly occurs when the Fermi level is placed within a spectral gap of the Hartree-shifted single-particle effective electronic Hamiltonian at these $U/t$ values. To address our assertion, we compute the single-particle density of states (DoS) near the Fermi level, which is given by $\sum_i\rho_e(i,\omega=0)$ as a function of $U/t$. Fig.\ref{fig: dome} (b) shows DoS as a function of $U/t$. Here, we multiply $U/t$ to capture the average interaction strength. Remarkably, the $U/t$ values where the maximum local critical temperature drops to zero in Fig.\ref{fig: dome}(a) roughly align with significant drops in the DoS as the Fermi level enters a single-particle gap\cite{supple}. Since the single-particle spectrum is fragmented with multiple spectral gaps due to the aperiodicity of quasicrystals, the inhomogeneous Hartree shift induces nontrivial changes in the spectrum as $U/t$ varies. Particularly, this allows a stronger attractive interaction to place the Fermi level within an effective single-particle gap, causing the abrupt disappearance of superconductivity.

Lastly, we note that for general values of $U/t$, the maximum and minimum values of the local critical temperature do not coincide. Hence, the dodecagonal quasicrystal exhibits not only a fully superconducting phase (green area in Fig.\ref{fig: dome} (a)) but also pattern-selective mixed phases (yellow area in Fig.\ref{fig: dome} (a)) in a broad range of attractive interaction strength, even without external fields or other pairing channels.

\textit{\tcr{Conclusion---}} In summary, we explored superconductivity in dodecagonal quasicrystals using the attractive on-site Hubbard model within the Bogoliubov–de Gennes framework. We revealed two distinct superconducting mechanisms in quasiperiodic systems: pattern-selective and interaction-tuned. The pattern-selective mechanism arises from inhomogeneous local electron density of states and diversity of local kinetic energy in terms of various coordination numbers and atomic distances, leading to spatially varying gaps and critical temperatures. This gives rise to mixed phases where superconducting and metallic regions coexist without external fields or unconventional pairing channels, with phase transitions observable in heat capacity beyond the BCS scenario. While, interaction-tuned superconductivity arises from the fragmentation of the electron spectrum, a hallmark of strong quasiperiodicity, where an inhomogeneous Hartree shift can align the Fermi level with a spectral gap, causing anomalous suppression at higher interaction strengths. Interaction-tuned pattern-selective superconductivity is not limited to specific tiling patterns but can emerge in general aperiodic systems, where inhomogeneous charge distribution and the diversified competition between interaction and kinetic terms are observed due to various local patterns. Our findings provide new insights into how strongly inhomogeneous quasiperiodic structures influence superconductivity and uncover unconventional critical phenomena in complex quasiperiodic systems.

Extending the mean-field framework to advanced many-body methods such as  Variational Monte Carlo (VMC)\cite{rubenstein2017introduction,yang2020deep,yamaji1998variational,yokoyama1988variational} and Dynamical Mean-Field Theory (DMFT)\cite{PhysRevB.72.155105,PhysRevB.95.024509,PhysRevB.105.205138}, to capture beyond mean-field correlations and gain deeper insights into electron-electron interactions in quasicrystalline superconductivity, is a potential future work. 
Additionally, exploring different tiling patterns, disorder effects, multi-orbital interactions, unconventional pairing mechanism and magnetic field effects could further reveal the exotic nature of superconductivity in quasicrystals\cite{kellendonk2015mathematics,cao2020kohn,liu2024nematic,sakai2019exotic,terashima2024anomalous}.

\section*{Acknowledgments}

J.M.J and S.B.L. were supported by National Research Foundation Grant (2021R1A2C109306013) and Nano Material Technology Development Program through the National Research Foundation of Korea(NRF) funded by Ministry of Science and ICT (RS-2023-00281839).

\bibliography{my1}

\begin{thebibliography}{10}

\bibitem{tinkham2004introduction}
Michael Tinkham.
\newblock {\em Introduction to superconductivity}.
\newblock Courier Corporation, 2004.

\bibitem{kamiya2018discovery}
Kazuya Kamiya, T~Takeuchi, N~Kabeya, N~Wada, T~Ishimasa, A~Ochiai, K~Deguchi,
  K~Imura, and NK~Sato.
\newblock Discovery of superconductivity in quasicrystal.
\newblock {\em Nature communications}, 9(1):154, 2018.

\bibitem{uri2023superconductivity}
Aviram Uri, Sergio~C de~la Barrera, Mallika~T Randeria, Daniel Rodan-Legrain,
  Trithep Devakul, Philip~JD Crowley, Nisarga Paul, Kenji Watanabe, Takashi
  Taniguchi, Ron Lifshitz, et~al.
\newblock Superconductivity and strong interactions in a tunable moir{\'e}
  quasicrystal.
\newblock {\em Nature}, 620(7975):762--767, 2023.

\bibitem{takemori2024superconducting}
Nayuta Takemori.
\newblock Superconducting quasicrystals.
\newblock {\em Israel Journal of Chemistry}, 64(10-11):e202300124, 2024.

\bibitem{sordelet1997quasicrystals}
Daniel~J Sordelet and Jean~Marie Dubois.
\newblock Quasicrystals perspectives and potential applications.
\newblock {\em MRS bulletin}, 22(11):34--39, 1997.

\bibitem{macia2020quasicrystals}
Enrique Maci{\'a}-Barber.
\newblock {\em Quasicrystals: fundamentals and applications}.
\newblock CRC Press, 2020.

\bibitem{suck2013quasicrystals}
J-B Suck, Michael Schreiber, and Peter H{\"a}ussler.
\newblock {\em Quasicrystals: An introduction to structure, physical properties
  and applications}, volume~55.
\newblock Springer Science \& Business Media, 2013.

\bibitem{sun2024enhancement}
Meng Sun, Tilen {\v{C}}ade{\v{z}}, Igor Yurkevich, and Alexei Andreanov.
\newblock Enhancement of superconductivity in the fibonacci chain.
\newblock {\em Physical Review B}, 109(13):134504, 2024.

\bibitem{duncan2024critical}
Callum~W Duncan.
\newblock Critical states and anomalous mobility edges in two-dimensional
  diagonal quasicrystals.
\newblock {\em Physical Review B}, 109(1):014210, 2024.

\bibitem{naka2005critical}
Michihiro Naka, Kazusumi Ino, and Mahito Kohmoto.
\newblock Critical level statistics of the fibonacci model.
\newblock {\em Physical Review B—Condensed Matter and Materials Physics},
  71(24):245120, 2005.

\bibitem{poon1992electronic}
2\_SJ Poon.
\newblock Electronic properties of quasicrystals an experimental review.
\newblock {\em Advances in Physics}, 41(4):303--363, 1992.

\bibitem{kohmoto1987critical}
Mahito Kohmoto, Bill Sutherland, and Chao Tang.
\newblock Critical wave functions and a cantor-set spectrum of a
  one-dimensional quasicrystal model.
\newblock {\em Physical Review B}, 35(3):1020, 1987.

\bibitem{jeon2021topological}
Junmo Jeon and SungBin Lee.
\newblock Topological critical states and anomalous electronic transmittance in
  one-dimensional quasicrystals.
\newblock {\em Physical Review Research}, 3(1):013168, 2021.

\bibitem{jagannathan2021fibonacci}
Anuradha Jagannathan.
\newblock The fibonacci quasicrystal: Case study of hidden dimensions and
  multifractality.
\newblock {\em Reviews of Modern Physics}, 93(4):045001, 2021.

\bibitem{xia2025superconductivity}
Yiyu Xia, Zhongdong Han, Kenji Watanabe, Takashi Taniguchi, Jie Shan, and
  Kin~Fai Mak.
\newblock Superconductivity in twisted bilayer wse2.
\newblock {\em Nature}, 637(8047):833--838, 2025.

\bibitem{wu2018theory}
Fengcheng Wu, Allan~H MacDonald, and Ivar Martin.
\newblock Theory of phonon-mediated superconductivity in twisted bilayer
  graphene.
\newblock {\em Physical review letters}, 121(25):257001, 2018.

\bibitem{chen2019signatures}
Guorui Chen, Aaron~L Sharpe, Patrick Gallagher, Ilan~T Rosen, Eli~J Fox, Lili
  Jiang, Bosai Lyu, Hongyuan Li, Kenji Watanabe, Takashi Taniguchi, et~al.
\newblock Signatures of tunable superconductivity in a trilayer graphene
  moir{\'e} superlattice.
\newblock {\em Nature}, 572(7768):215--219, 2019.

\bibitem{torma2022superconductivity}
P{\"a}ivi T{\"o}rm{\"a}, Sebastiano Peotta, and Bogdan~A Bernevig.
\newblock Superconductivity, superfluidity and quantum geometry in twisted
  multilayer systems.
\newblock {\em Nature Reviews Physics}, 4(8):528--542, 2022.

\bibitem{emery2000charge}
VJ~Emery and SA~Kivelson.
\newblock Charge inhomogeneity and high temperature superconductivity.
\newblock {\em Journal of Physics and Chemistry of Solids}, 61(3):467--471,
  2000.

\bibitem{ghosal2001inhomogeneous}
Amit Ghosal, Mohit Randeria, and Nandini Trivedi.
\newblock Inhomogeneous pairing in highly disordered s-wave superconductors.
\newblock {\em Physical Review B}, 65(1):014501, 2001.

\bibitem{yoshii2025brillouin}
Mao Yoshii, Sota Kitamura, and Takahiro Morimoto.
\newblock Brillouin zone folding method for quasiperiodic superconductivity in
  multilayer systems: Application to electronic structure and optical
  responses.
\newblock {\em Physical Review B}, 111(2):024206, 2025.

\bibitem{balents2020superconductivity}
Leon Balents, Cory~R Dean, Dmitri~K Efetov, and Andrea~F Young.
\newblock Superconductivity and strong correlations in moir{\'e} flat bands.
\newblock {\em Nature Physics}, 16(7):725--733, 2020.

\bibitem{mace2017critical}
Nicolas Mac{\'e}, Anuradha Jagannathan, Pavel Kalugin, R{\'e}my Mosseri, and
  Fr{\'e}d{\'e}ric Pi{\'e}chon.
\newblock Critical eigenstates and their properties in one-and two-dimensional
  quasicrystals.
\newblock {\em Physical Review B}, 96(4):045138, 2017.

\bibitem{takemori2020physical}
Nayuta Takemori, Ryotaro Arita, and Shiro Sakai.
\newblock Physical properties of weak-coupling quasiperiodic superconductors.
\newblock {\em Physical Review B}, 102(11):115108, 2020.

\bibitem{tokumoto2024superconductivity}
Yuki Tokumoto, Kotaro Hamano, Sunao Nakagawa, Yasushi Kamimura, Shintaro
  Suzuki, Ryuji Tamura, and Keiichi Edagawa.
\newblock Superconductivity in a van der waals layered quasicrystal.
\newblock {\em Nature communications}, 15(1):1529, 2024.

\bibitem{liu2024nematic}
Yu-Bo Liu, Jing Zhou, and Fan Yang.
\newblock Nematic superconductivity and its critical vestigial phases in the
  quasicrystal.
\newblock {\em Physical Review Letters}, 133(13):136002, 2024.

\bibitem{verbin2013observation}
Mor Verbin, Oded Zilberberg, Yaacov~E Kraus, Yoav Lahini, and Yaron Silberberg.
\newblock Observation of topological phase transitions in photonic
  quasicrystals.
\newblock {\em Physical review letters}, 110(7):076403, 2013.

\bibitem{ghadimi2021topological}
Rasoul Ghadimi, Takanori Sugimoto, K~Tanaka, and Takami Tohyama.
\newblock Topological superconductivity in quasicrystals.
\newblock {\em Physical Review B}, 104(14):144511, 2021.

\bibitem{fan2021enhanced}
Zhijie Fan, Gia-Wei Chern, and Shi-Zeng Lin.
\newblock Enhanced superconductivity in quasiperiodic crystals.
\newblock {\em Physical Review Research}, 3(2):023195, 2021.

\bibitem{cain2020layer}
Jeffrey~D Cain, Amin Azizi, Matthias Conrad, Sin{\'e}ad~M Griffin, and Alex
  Zettl.
\newblock Layer-dependent topological phase in a two-dimensional quasicrystal
  and approximant.
\newblock {\em Proceedings of the National Academy of Sciences},
  117(42):26135--26140, 2020.

\bibitem{terashima2024anomalous}
Taichi Terashima, Yuki Tokumoto, Kotaro Hamano, Takako Konoike, Naoki Kikugawa,
  and Keiichi Edagawa.
\newblock Anomalous upper critical field in the quasicrystal superconductor
  ta1. 6te.
\newblock {\em npj Quantum Materials}, 9(1):56, 2024.

\bibitem{araujo2019conventional}
Ronaldo~N Ara{\'u}jo and Eric~C Andrade.
\newblock Conventional superconductivity in quasicrystals.
\newblock {\em Physical Review B}, 100(1):014510, 2019.

\bibitem{sakai2017superconductivity}
Shiro Sakai, Nayuta Takemori, Akihisa Koga, and Ryotaro Arita.
\newblock Superconductivity on a quasiperiodic lattice: Extended-to-localized
  crossover of cooper pairs.
\newblock {\em Physical Review B}, 95(2):024509, 2017.

\bibitem{sakai2019exotic}
Shiro Sakai and Ryotaro Arita.
\newblock Exotic pairing state in quasicrystalline superconductors under a
  magnetic field.
\newblock {\em Physical Review Research}, 1(2):022002, 2019.

\bibitem{ghadimi2024quasiperiodic}
Rasoul Ghadimi and Bohm-Jung Yang.
\newblock Quasiperiodic pairing in graphene quasicrystals.
\newblock {\em Nano Letters}, 2024.

\bibitem{kellendonk2015mathematics}
Johannes Kellendonk, Daniel Lenz, and Jean Savinien.
\newblock {\em Mathematics of aperiodic order}, volume 309.
\newblock Springer, 2015.

\bibitem{yamada2022four}
Tsunetomo Yamada.
\newblock A four-dimensional model for the ba--ti--o dodecagonal quasicrystal.
\newblock {\em Structural Science}, 78(2):247--252, 2022.

\bibitem{gahler1988crystallography}
Franz G{\"a}hler.
\newblock Crystallography of dodecagonal quasicrystals.
\newblock in Quasicrystalline Materials, edited by C. Janot and J.M. Dubois
  (World Scientific, Singapore, 1988), pp. 272–284.

\bibitem{supple}
{See Supplemental Material at [URL will be inserted by publisher] for detailed
  information of construction and structure of the dodecagonal quasicrystal,
  size of Cooper pair and coherence length, local order parameters as the
  function of temperature and effective single particle spectrum around the
  Fermi level.}

\bibitem{zhu2016bogoliubov}
Jian-Xin Zhu.
\newblock {\em Bogoliubov-de Gennes method and its applications}, volume 924.
\newblock Springer, 2016.

\bibitem{collings2013applied}
EW~Collings.
\newblock {\em Applied Superconductivity, Metallurgy, and Physics of Titanium
  Alloys: Fundamentals Alloy Superconductors: Their Metallurgical, Physical,
  and Magnetic-Mixed-State Properties}.
\newblock Springer Science \& Business Media, 2013.

\bibitem{sharma2015type}
RG~Sharma and RG~Sharma.
\newblock Type ii superconductors.
\newblock {\em Superconductivity: Basics and Applications to Magnets}, pages
  49--69, 2015.

\bibitem{zeng2016generalized}
Qi-Bo Zeng, Shu Chen, and Rong L{\"u}.
\newblock Generalized aubry-andr{\'e}-harper model with p-wave superconducting
  pairing.
\newblock {\em Physical Review B}, 94(12):125408, 2016.

\bibitem{cao2020kohn}
Ye~Cao, Yongyou Zhang, Yu-Bo Liu, Cheng-Cheng Liu, Wei-Qiang Chen, and Fan
  Yang.
\newblock Kohn-luttinger mechanism driven exotic topological superconductivity
  on the penrose lattice.
\newblock {\em Physical Review Letters}, 125(1):017002, 2020.

\bibitem{li1993mixed}
QP~Li, BEC Koltenbah, and Robert Joynt.
\newblock Mixed s-wave and d-wave superconductivity in high-t c systems.
\newblock {\em Physical Review B}, 48(1):437, 1993.

\bibitem{o1995s}
C~O’Donovan and JP~Carbotte.
\newblock s-and d-wave mixing in high-t c superconductors.
\newblock {\em Physical Review B}, 52(22):16208, 1995.

\bibitem{rubenstein2017introduction}
Brenda Rubenstein.
\newblock Introduction to the variational monte carlo method in quantum
  chemistry and physics.
\newblock {\em Variational Methods in Molecular Modeling}, pages 285--313,
  2017.

\bibitem{yang2020deep}
Li~Yang, Zhaoqi Leng, Guangyuan Yu, Ankit Patel, Wen-Jun Hu, and Han Pu.
\newblock Deep learning-enhanced variational monte carlo method for quantum
  many-body physics.
\newblock {\em Physical Review Research}, 2(1):012039, 2020.

\bibitem{yamaji1998variational}
Kunihiko Yamaji, Takaashi Yanagisawa, Takeshi Nakanishi, and Soh Koike.
\newblock Variational monte carlo study on the superconductivity in the
  two-dimensional hubbard model.
\newblock {\em Physica C: Superconductivity}, 304(3-4):225--238, 1998.

\bibitem{yokoyama1988variational}
Hisatoshi Yokoyama and Hiroyuki Shiba.
\newblock Variational monte-carlo studies of superconductivity in strongly
  correlated electron systems.
\newblock {\em Journal of the Physical Society of Japan}, 57(7):2482--2493,
  1988.

\bibitem{PhysRevB.72.155105}
M.~V. Sadovskii, I.~A. Nekrasov, E.~Z. Kuchinskii, Th. Pruschke, and V.~I.
  Anisimov.
\newblock Pseudogaps in strongly correlated metals: A generalized dynamical
  mean-field theory approach.
\newblock {\em Phys. Rev. B}, 72:155105, Oct 2005.

\bibitem{PhysRevB.95.024509}
Shiro Sakai, Nayuta Takemori, Akihisa Koga, and Ryotaro Arita.
\newblock Superconductivity on a quasiperiodic lattice: Extended-to-localized
  crossover of cooper pairs.
\newblock {\em Phys. Rev. B}, 95:024509, Jan 2017.

\bibitem{PhysRevB.105.205138}
Shiro Sakai and Nayuta Takemori.
\newblock Doped mott insulator on a penrose tiling.
\newblock {\em Phys. Rev. B}, 105:205138, May 2022.

\end{thebibliography}
\bibliographystyle{unsrt}

\newpage

\renewcommand{\thesection}{\arabic{section}}
\setcounter{section}{0}
\renewcommand{\thefigure}{S\arabic{figure}}
\setcounter{figure}{0}
\renewcommand{\theequation}{S\arabic{equation}}
\setcounter{equation}{0}

\begin{widetext}
	\section*{Supplemental Material of Interaction tuned pattern-selective superconductivity: Application to the dodecagonal quasicrystal}
\section{Construction and structure of the dodecagonal quasicrystal}
    \label{sec:1}
    In this section, we provide a detail of the cut-and-project construction of the dodecagonal quasicrystal. The cut-and-project method is a well-established technique for constructing quasicrystals, which exhibit long-range order without periodicity. Unlike periodic crystals, which are defined by translational symmetry in a single-dimensional lattice, quasicrystals can be understood as projections of higher-dimensional lattices onto a lower-dimensional physical space. This method explains the origin of quasiperiodicity and the presence of non-crystallographic rotational symmetries, especially twelvefold symmetry, which we will particularly discuss here.

    Starting from the standard 4D hypercubic lattice, let us define the projection maps $M$ and $M^\perp$ to the physical space and its orthogonal complement space, respectively as below
    \begin{align}
\label{projection_ddQC}
M = \frac{1}{a\sqrt{2}}
\begin{bmatrix}
\cos\left(\frac{\pi}{6}\right) & 1 & 0 & \cos\left(\frac{2\pi}{3}\right) \\
-\sin\left(\frac{\pi}{6}\right) & 0 & 1 & \sin\left(\frac{2\pi}{3}\right),
\end{bmatrix}
\end{align}
and
\begin{align}
\label{projection_ddQC}
M^{\perp} = \frac{1}{a\sqrt{2}}
\begin{bmatrix}
-\cos\left(\frac{\pi}{6}\right) & 1 & 0 & \cos\left(\frac{2\pi}{3}\right) \\
-\sin\left(\frac{\pi}{6}\right) & 0 & 1 & -\sin\left(\frac{2\pi}{3}\right),
\end{bmatrix}
\end{align}
Here, $a=1.366$. Since these projections are incommensurate, we should set a compact window $K$ so that among the hypercubic lattice points, only those whose images under $M^\perp$ fall within $K$ are selectively projected onto the physical space according to $M$. To construct dodecagonal symmetric structure, we choose the dodecagonal shaped window. In detail, we first define two vectors, $\vec{P}_1$ and $\vec{P}_2$ on the perpendicular space, which are given by
\begin{align}
    \vec{P}_1=M^\perp\begin{pmatrix}0\\1/2\\0\\0\end{pmatrix} \text{ and } \vec{P}_2=M^\perp\begin{pmatrix}-\sqrt{3}/6\\\sqrt{3}/6\\0\\0\end{pmatrix}.
\end{align}
Now we consider their twelvefold rotation copies with respect to the origin. Specifically, we consider the set of position vectors, $\{R^n\vec{P}_i \ \ \vert \ \ 0\le n\le11, i=1,2\}$, where $R=\begin{pmatrix} \cos{\pi/6} & \sin{\pi/6} \\ -\sin{\pi/6} & \cos\pi/6\end{pmatrix}$. This forms the dodecagonal star-shaped window $K$ illustrated in Fig.1(a) in the main text. Then, the set of dodecagonal quasicrystalline lattice points, $QL$ is given by $QL=\{M\vec{x} \ \ \vert \ \ M^\perp\vec{x}\in K  \}$. Here, $\vec{x}$ is the hypercubic lattice point.

To check the dodecagonal structure of the resulting quasicrystalline lattice, we calculate the Bragg peaks assuming the identical atoms placed on each lattice points in $QL$. Remind that the structure factor is given by
\begin{align}
    \label{structurefactor}
    &S(\vec{k})=\sum_i f_i e^{i\vec{k}\cdot \vec{r}_i}.
\end{align}
Here, $\vec{k}$ is the momentum vector and $\vec{r}_i$ is the quasicrystalline lattice point. $f_i$ is the atomic form factor, which we let $f_i=1$ for simplicity. Then, the Bragg peaks intensity is given by $I(\vec{k})=\vert S(\vec{k})\vert^2$. Fig.\ref{fig: bragg} shows the Bragg peak intensity of dodecagonal quasicrystal. It reveals 12-fold rotational symmetric structure. Note that some extra peaks appear due to the finite size effect.
\begin{figure}[h]
    \centering
\includegraphics[width=0.6\linewidth]{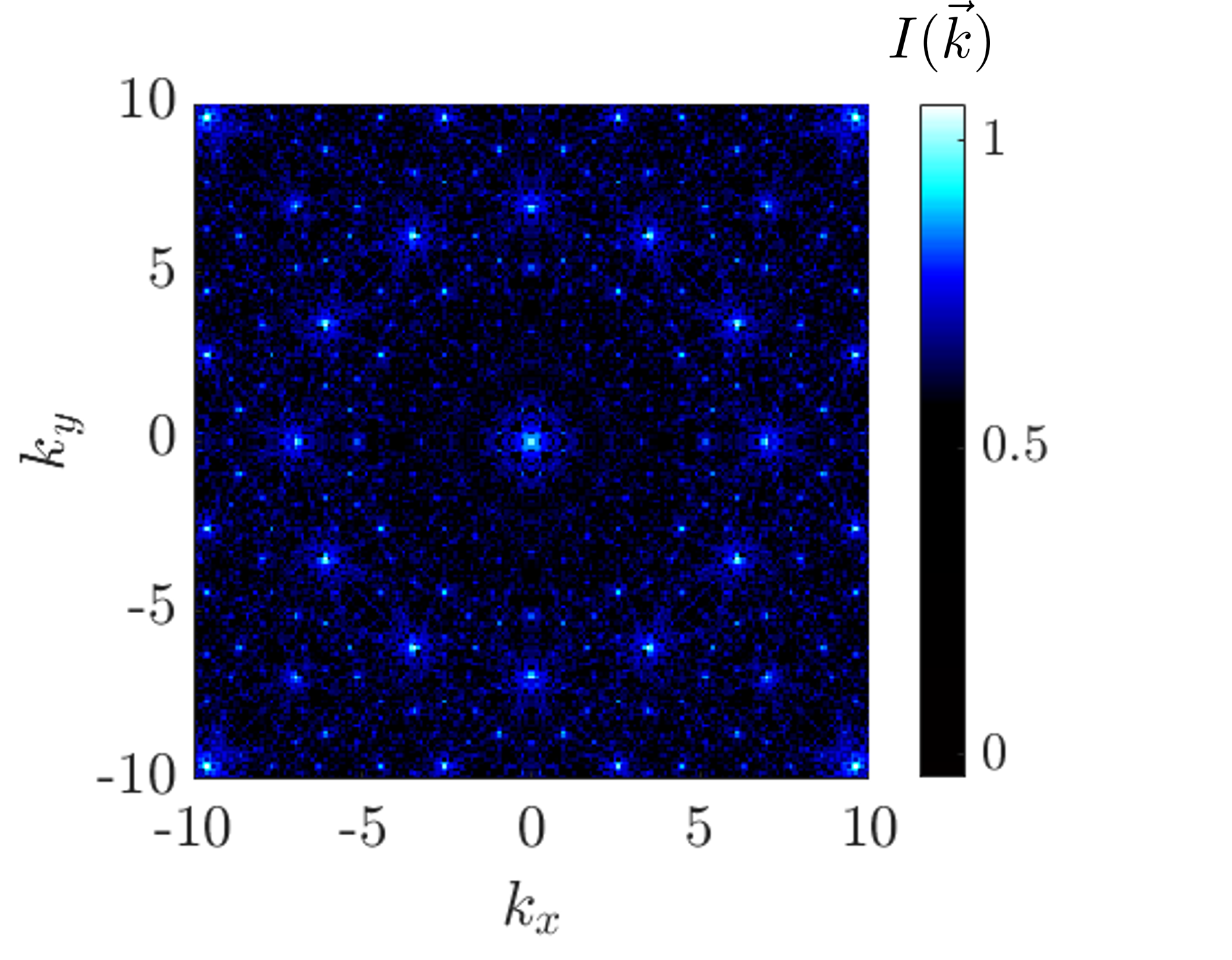}
    \caption{\label{fig: bragg} Normalized log scale Bragg peaks of the dodecagonal quasicrystal constructed by the cut-and-project scheme. The number of sites is 5437. 12-fold symmetric pattern of Bragg peaks is originated from the dodecagonal symmetric long-range order of quasicrystal.}
    \label{fig:bragg}
\end{figure}

\section{Size of Cooper pair and Coherence length}
\label{sec:2} In this section, we investigate the average size of the Cooper pair and the coherence length to show that the finite-size effect can be excluded for an attractive interaction around $U/t=1$, given the system size considered in the main text. To this end, we compute the off-site pairing correlation, $\Delta_{ij}$ given by
\begin{align}
    \label{coopersize}
    &\Delta_{ij}=\frac{1}{2}\left(\braket{c_{i\uparrow}c_{j\downarrow}}-\braket{c_{i\downarrow}c_{j\uparrow}}\right)
\end{align}
In terms of Bogoliubov transformation, this can be rewritten as
\begin{align}
    \label{coopersize2}
    &\Delta_{ij}=\frac{1}{4}\sum_n(u_{i\uparrow}^n v_{j\downarrow}^{n*}+u_{j\uparrow}^n v_{i\downarrow}^{n*})\tanh{\frac{E_n}{2T}}
\end{align}
By examining scaling behavior of $\Delta_{ij}$ for long distances, we can get the coherence length. Specifically, if $\Delta_{ij}\sim e^{-R_{ij}/\xi}$, then $\xi$ is the coherence length which represents the size of Cooper pair. When the coherence length is much smaller than the system size, we can exclude the finite size effect. Since the non-uniform local critical temperature is observed even at zero temperature, it is sufficient to explore the case of $T=0$ to show our assertion is not a simple finite size effect.

Fig.\ref{fig:coherence} illustrates $\Delta_{ij}$ as a function of $R_{ij}$ at zero temperature. The Cooper pair is exponentially localized with coherence length $\xi\approx 3.912\ll R_c$, where $R_c=40$ is the radius of the system we consider.
\begin{figure}[h]
    \centering
\includegraphics[width=0.6\linewidth]{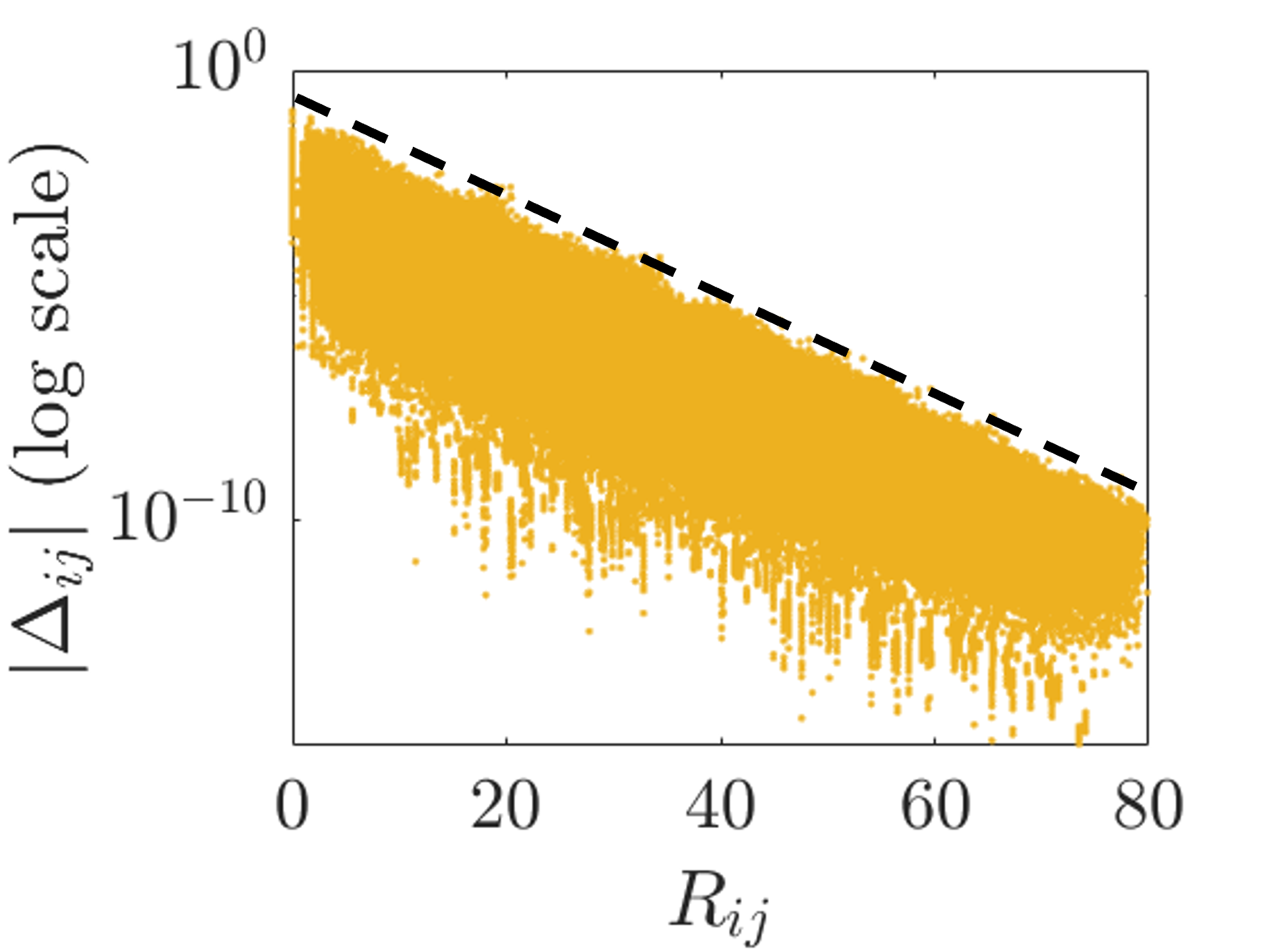}
    \caption{\label{fig: coherence} Off-site pairing correlation (log-scale) as a function of distance. By fitting $\Delta_{ij}\sim e^{-R_{ij}/\xi}$, the coherence length, $\xi\approx 3.912$. The number of sites is 5437. The radius of the system is $40$, which is much larger than the coherence length. $T=0$, $U/t=1$.
    \label{fig:coherence}}
\end{figure}

\section{Local order parameters as the function of temperature}

In this section, we explore temperature dependence of local order parameters, $\Delta_i$. Fig.\ref{fig: order} shows $\Delta_i$ as the function of temperature. Note that each local order parameter roughly follows $\Delta_i=\Delta_0 \sqrt{1-\frac{T}{T_{c,i}}}$ for $T\le T_{c,i}$ where $T_{c,i}$ is the local critical temperature. The different local critical temperature leads to the different curvatures even for the same $\Delta_0$ in Fig.\ref{fig: order}. Additionally, the minimum $\Delta_i$ is zero at $T=0$ for $U/t=1$. This indicates the emergence of mixed phase even at zero temperature as we have shown in Fig.2 in the main text.
\begin{figure}[h]
    \centering
\includegraphics[width=0.8\linewidth]{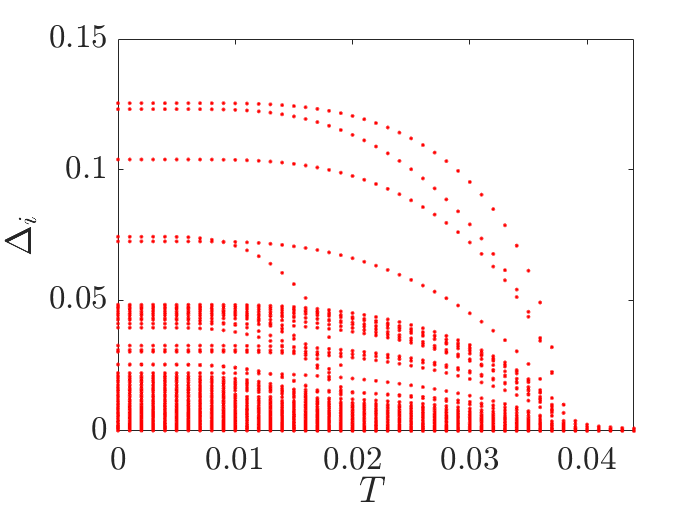}
    \caption{\label{fig: order} Local order parameters as the functions of temperature. The number of sites is 5437. $t=1$. $U/t=1$.}
    \label{fig:energy}
\end{figure}

\section{Effective single particle spectrum around the Fermi level}
    \label{sec:3} In this section, to unveil the underlying mechanism of the anomalous $U$-dependent phase transitions of superconductivity in dodecagonal quasicrystal, we investigate the effective single particle spectrum around the Fermi level ($E_F=0$) of the Hartree shifted effective electronic Hamiltonian, $\mathcal{H}_{i\sigma,j\sigma}$, which is given by
    \begin{align}
        \label{effectiveH}
        &\mathcal{H}_{i\sigma,j\sigma}=-t_{ij}-\tilde{\mu}\delta_{ij}.
    \end{align}
Here, the hopping integrals $t_{ij}$ is given by
\begin{align}
    \label{tij}
    &t_{ij}=\begin{cases} te^{1-R_{ij}} \ \ ,0<R_{ij}\le 1 \\ 0 \ \ \ \ \ \ \ \ \ \ \ \ \mathrm{,otherwise} \end{cases},
\end{align}
where $R_{ij}$ is the distance between two sites $i$ and $j$. $\tilde{\mu}=\mu+Un_{i\bar{\sigma}}$, where $n_{i\bar{\sigma}}$ is the local charge density of electron with opposite spin. When the external magnetic field is absent, $n_{i\uparrow}=n_{i\downarrow}=n_{i}/2$. Thus, in our case, $\tilde{\mu}=\mu+Un_{i}/2$. The second term of $\tilde{\mu}$, which is called the Hartree shift gives rise to the anomalous $U$-dependence of superconductivity by changing the characteristics of $\mathcal{H}$. We maintain $\mu$, external chemical potential by using the particle reservoir. Then, the Fermi level of the electron at the zero temperature is simply zero for $\mathcal{H}$. However, due to the Hartree shift term, the characteristics near the Fermi level would be drastically changed as the function of $U$.

Fig.\ref{fig:energy} shows the spectrum of $\mathcal{H}$ near zero energy (highlighted by red dashed lines) for three $U/t$ values. For $U/t = 0.75$ and $U/t = 1.25$, the Fermi level lies within the spectral gap, making the system effectively insulating and prohibiting superconductivity. In contrast, for $U/t = 1$, the Fermi level corresponds to a metallic regime, allowing superconductivity. This leads to the abrupt emergence and disappearance of superconductivity with increasing $U/t$, resulting in anomalous phase transitions in the dodecagonal quasicrystal.
\begin{figure}[h]
    \centering
\includegraphics[width=0.9\linewidth]{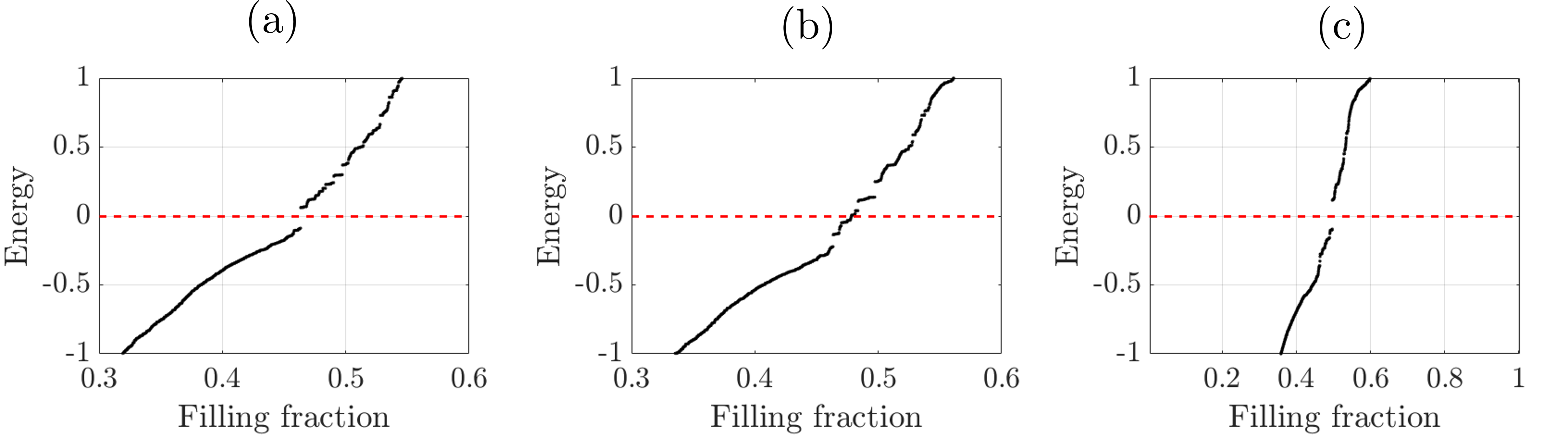}
    \caption{\label{fig: energy} Effective Hartree shifted electronic spectrum near the zero energy for $U/t=$ (a) 0.75 (b) 1 and (c) 1.25, respectively. The red dashed lines are drawn for emphasizing the position of the Fermi level. External chemical potential is $\mu=-0.25t$, corresponding to the nearly half-filled Fermi level of non-interacting electronic Hamiltonian. (a,c) The Fermi level is placed within the spectral gap of the effective single particle spectrum. Thus, the superconductivity disappears at these values of $U/t$. (b) The Fermi level is placed as the metallic regime. This leads to the formation of the stable superconductivity. The number of sites is 5437. $t=1$. Compare the results shown in Fig.3 in the main text.}
    \label{fig:energy}
\end{figure}

\end{widetext}

\end{document}